\documentclass[prd,onecolumn]{revtex4}
\usepackage{dcolumn}
\usepackage{multirow}
\usepackage{graphicx}
\usepackage{amssymb}
\usepackage{bm}
\usepackage{hyperref}
\usepackage{epstopdf}
\usepackage{color}
\usepackage{mathrsfs}
\usepackage{amsmath,amssymb,amsthm}
\usepackage{rotating}
\usepackage{sverb, longtable}
\begin{document}
\title{Dark Energy Constraints in light of Pantheon SNe Ia, BAO, Cosmic Chronometers and CMB Polarization and Lensing Data}
\author{Deng Wang}
\email{cstar@mail.nankai.edu.cn}
\affiliation{Theoretical Physics Division, Chern Institute of Mathematics, Nankai University, Tianjin 300071, China}

\begin{abstract}
To explore whether there is new physics going beyond the standard cosmological model or not, we constrain seven cosmological models by combining the latest and largest Pantheon Type Ia supernovae sample with the data combination of baryonic acoustic oscillations, cosmic microwave background radiation, Planck lensing and cosmic chronometers. We find that a spatially flat universe is preferred in the framework of $\Lambda$CDM cosmology, that the constrained equation of state of dark energy is very consistent with the cosmological constant hypothesis in the $\omega$CDM model, that there is no evidence of dynamical dark energy in the dark energy density-parametrization model, that there is no hint of interaction between dark matter and dark energy in the dark sector of the universe in the decaying vacuum model, and that there does not exist the sterile neutrino in the neutrino sector of the universe in the $\Lambda$CDM model. We also give the 95$\%$ upper limit of the total mass of three active neutrinos $\Sigma m_\nu<0.178$ eV under the assumption of $\Lambda$CDM scenario. It is clear that there is no any departure from the standard cosmological model based on current observational datasets.

\end{abstract}
\maketitle

\section{Introduction}
With gradually mounting astronomical data and unceasingly improved statistical techniques, it is very promising for human beings to unveil the mysterious phenomena at cosmological scale in the near future. The main challenge of modern cosmology is to understand the nature of cosmic late-time acceleration, which has been confirmed by a large number of observations such as measurements of Type Ia supernovae (SNe Ia) distances \cite{1,2}, peaks of baryon acoustic oscillations (BAO) in the large-scale correlation function of galaxies \cite{3}, and the power spectrum of fluctuations in the cosmic microwave background (CMB) \cite{4,5}. Most recently, on August 17, 2017 at 12:41:04 UTC, the first multimessenger gravitational-wave observation of a binary neutron star inspiral made by LIGO-Virgo detector network, once again, gives a convincing evidence that the universe is undergoing a phase of accelerated expansion \cite{6,7}. Up to now, to describe this accelerated mechanism, theoretical physicists propose two main classes of models \cite{8}, i.e., the so-called dark energy (DE) and modified gravities. The former maintains the correctness of the general relativity (GR) and introduces an exotic matter source in the Einstein equation, while the latter modifies the standard lagrangian of GR based on some reasonable physical consideration.

At present, in light of rich observed data, we just know several primary properties of DE: (i) a cosmic fluid with an equation of state (EoS) $\omega\approx-1$, which violates the strong energy condition; (ii) the DE fluid is homogeneously permeated in the universe and has no the property of clustering unlike the dark matter (DM). To characterize the DE phenomena, the simplest candidate is the standard cosmological model, i.e., the cosmological constant ($\Lambda$) plus cold DM ($\Lambda$CDM) scenario. This model can successfully explain various observations at both large and small scales. However, it faces at least two intractable problems, namely the coincidence and fine-tuning problems \cite{9}. This indicates that the $\Lambda$CDM model may be not the underlying one governing the background evolution and large scale structure formation of the universe, and that it should be extended to a more complicated one or even replaced with another physical-driven scenario. Since the accelerated universe is discovered, there have been a great deal of DE models proposed by theoreticians to overcome the difficulties that $\Lambda$CDM meets in the last twenty years. With the endless data coming, it is more important for cosmologists to test the validity of some known and simple models rather than to develop new theoretical paradigms.

SNe Ia is a powerful geometrical distance indicator to probe the expansion history of the universe, particularly, the EoS of DE. During the past three years, the standard SNe Ia sample to constrain the cosmological models is the `` Joint Light-curve Analysis '' (JLA) one constructed from Supernova Legacy Survey (SNLS) and Sloan Digital Sky Survey (SDSS), which consists of 740 data points covering the redshift range $0.01<z<1.3$ \cite{10}. Most Recently, D. Scolnic {\it et al.} have reported a larger `` Pantheon '' SNe Ia sample than JLA one by combining the subset of 276 new Pan-STARRS1 (PS1) SNe Ia with useful distance estimates of SNe Ia from SNLS, SDSS, low-$z$ and Hubble space telescope (HST) samples \cite{11}. This sample consists of 1049 spectroscopically confirmed SNe Ia covering the redshift range $0.01<z<2.3$. In this work, our basic idea is to implement the DE constraints by using the latest and largest Pantheon SNe Ia sample to date. We obtain strict constraints on the parameters of seven cosmological models by combining the Pantheon data with the other four datasets.

This work is organized as follows. In the next section, we introduce seven cosmological models used in this analysis. In Section III, we describe the latest observational datasets and our analysis methodology. In Section IV, we exhibit the numerical analysis results. The conclusions are presented in the final section.

\section{Seven cosmological models}
In this section, we introduce seven cosmological models to be constrained by using the astronomical datasets. Note that throughout this work, we investigate these seven models in a Friedmann-Robertson-Walker (FRW) universe in the framework of GR, and just concentrate on the late-time cosmology, consequently neglecting the contribution from radiation in the cosmic pie. Starting from the so-called Friedmann equations, we derive the corresponding dimensionless Hubble parameter (DHP) for the $\Lambda$CDM model as
\begin{equation}
E_1(z)=\left[\Omega_{m}(1+z)^3+1-\Omega_{m}\right]^{\frac{1}{2}}, \label{1}
\end{equation}
while for the non-flat $\Lambda$CDM (o$\Lambda$CDM) model it is expressed as
\begin{equation}
E_2(z)=\left[\Omega_{m}(1+z)^3+\Omega_{K}(1+z)^2+1-\Omega_{m}-\Omega_{K}\right]^{\frac{1}{2}}, \label{2}
\end{equation}
where $\Omega_{m}$ and $\Omega_{K}$ are present-day matter and curvature density parameters, respectively.

To achieve the goal of constraining the EoS of DE, we consider its simplest parametrization $\omega(z)=\omega$, i.e., the $\omega$CDM model, where the DE is a single negative pressure fluid. The DHP for the spatially flat $\omega$CDM model can be expressed as
\begin{equation}
E_3(z)=\left[\Omega_{m}(1+z)^3+(1-\Omega_{m})(1+z)^{3(1+\omega)}\right]^{\frac{1}{2}}. \label{3}
\end{equation}

For a long time, an important topic in the field of modern cosmology is whether the DE evolves over time or not. To address this issue, we consider a DE density-parametrization (DEDP) model recently proposed by us \cite{12}, and its DHP is shown as
\begin{equation}
E_4(z)=\left[\Omega_{m}(1+z)^{3}+(1-\Omega_{m})(1+\beta-\frac{\beta}{1+z})\right]^{\frac{1}{2}},   \label{4}
\end{equation}
where $\beta$ is the typical free parameter characterizing this parametrization model. It is easy to see that this model reduces to $\Lambda$CDM when $\beta=0$, and that if $\beta$ has any departure from zero, the DE will be dynamical.

Another important concern for cosmologists is whether or not there exists the interaction between DM and DE in the dark sector of the universe. To investigate this topic, we constrain an interesting decaying vacuum (DV) model proposed by Wang and Meng \cite{13}, and its DHP is written as
\begin{equation}
E_5(z)=\left[\frac{3\Omega_{m}}{3-\epsilon}(1+z)^{3-\epsilon}+1-\frac{3\Omega_{m}}{3-\epsilon}\right]^{\frac{1}{2}},   \label{5}
\end{equation}
where $\epsilon$ denotes a free parameter of this DV model. It is worth noting that $\epsilon$ means a small modified matter expansion rate. $\epsilon<0$ indicates that the momentum transfers from DM to DE and vice versa.

Based on our previous works \cite{a1,a2,a3}, it is interesting to explore the neutrino physics in the framework of standard cosmological model using the latest Pantheon data.  To weigh the total mass of three active neutrinos $\Sigma m_{\nu}$, we consider the $\Lambda$CDM plus a varying $\Sigma m_{\nu}$ model ($\Lambda\nu$) by keeping the effective number of relativistic species $N_{eff}=3.046$. In addition, we also explore the possibility of existence of massless sterile neutrinos by  assuming $\Sigma m_{\nu} = 0.06$ eV with a degenerate mass hierarchy. For simplicity, We refer to this model as `` $\Lambda$s '' hereafter.

Furthermore, we take the linear perturbations of background metric into account. As usual, the scalar mode perturbation of FRW spacetime is expressed as \cite{14,15,16}
\begin{equation}
ds^2=-(1+2\Phi)dt^2+2a\partial_iBdtdx+a^2[(1-2\Psi)\delta_{ij}+2\partial_i\partial_jE]dx^idx^j,  \label{6}
\end{equation}
where $\Phi$ and $\Psi$ are the linear gravitational potentials. By use of the synchronous gauge $\Psi=\eta$, $\Phi=B=0$ and $E=-(h+6\eta)/2k^2$, the energy-momentum conservation equations of the cosmic fluids are written as \cite{15}
\begin{equation}
\delta'=-3(\frac{\delta p}{\delta\rho}-\tilde{\omega})\mathcal{H}\delta-(1+\tilde{\omega})(\theta+\frac{h'}{2}), \label{7}
\end{equation}
\begin{equation}
\theta'=\frac{\delta p}{\delta\rho}\frac{k^2\delta}{1+\tilde{\omega}}+(3\tilde{\omega}-1)\mathcal{H}\theta-\frac{\tilde{\omega}'}{1+\tilde{\omega}'}\theta-k^2\delta, \label{8}
\end{equation}
where $\tilde{\omega}$, $\mathcal{H}$, $\sigma$,  $\theta$ and $\delta$ are, respectively, the EoSs of cosmic fluids, conformal HP, shear, velocity perturbation and density perturbation, and the prime denotes the derivative with respect to the conformal time. Subsequently, the DE perturbations can be expressed as
\begin{equation}
\delta_{de}'=3\mathcal{H}(\omega_{de}-c_s^2)\left[\delta_{de}+3\mathcal{H}(1+\omega_{de})\frac{\theta_{de}}{k^2}\right]-3\mathcal{H}\omega'_{de}\frac{\theta_{de}}{k^2}-(1+\omega_{de})(\theta_{de}+\frac{h'}{2}),   \label{9}
\end{equation}
\begin{equation}
\theta_{de}'=\frac{c_s^2}{1+\omega_{de}}k^2\delta_{de}+(3c_s^2-1)\mathcal{H}\theta_{de}, \label{10}
\end{equation}
where $\omega_{de}$ and $c_s^2$ are the effective EoS of DE and the physical sound speed in the rest frame, respectively. In order to avoid the unphysical sound speed, we have used $c_s^2=1$ in the following analysis. At the same time, to calculate more smoothly, we also take $\sigma=0$ numerically. Notice that the effective EoSs of DE of the $\Lambda$CDM, $o\Lambda$CDM, $\omega$CDM, DEDP, DV, $\Lambda\nu$ and $\Lambda$s models can be, respectively, expressed as $\omega_{de} =$ -1, -1, $\omega$, $-1+\frac{\beta}{3[(1+\beta)(1+z)-\beta]}$ \cite{12}, $-1+\frac{(1+z)^{3-\epsilon}-(1+z)^{3}}{\frac{3}{3-\epsilon}(1+z)^{3-\epsilon}-(1+z)^{3}+\frac{\tilde{\Omega_{\Lambda}}}{\Omega_{m}}}$ \cite{13}, -1 and -1, where $\tilde{\Omega_{\Lambda}}$ denotes the dimensionless ground state value of the vacuum.

\section{Observational datasets and analysis methodology}

In this section, we utilize the latest cosmological observations including the Pantheon SNe Ia sample to constrain the above models. The corresponding parameter spaces of these seven models can be shown as
\begin{equation}
\mathbf{P_1}=\{\Omega_bh^2, \quad \Omega_ch^2, \quad 100\theta_{MC}, \quad \tau, \quad  \mathrm{ln}(10^{10}A_s), \quad  n_s \},   \label{6}
\end{equation}
\begin{equation}
\mathbf{P_2}=\{\Omega_bh^2, \quad \Omega_ch^2, \quad 100\theta_{MC}, \quad \tau, \quad \Omega_K, \quad  \mathrm{ln}(10^{10}A_s), \quad  n_s \},   \label{7}
\end{equation}
\begin{equation}
\mathbf{P_3}=\{\Omega_bh^2, \quad \Omega_ch^2, \quad 100\theta_{MC}, \quad \tau, \quad \omega, \quad  \mathrm{ln}(10^{10}A_s), \quad  n_s \},   \label{8}
\end{equation}
\begin{equation}
\mathbf{P_4}=\{\Omega_bh^2, \quad \Omega_ch^2, \quad 100\theta_{MC}, \quad \tau, \quad \beta, \quad  \mathrm{ln}(10^{10}A_s), \quad  n_s \},   \label{9}
\end{equation}
\begin{equation}
\mathbf{P_5}=\{\Omega_bh^2, \quad \Omega_ch^2, \quad 100\theta_{MC}, \quad \tau, \quad \epsilon, \quad  \mathrm{ln}(10^{10}A_s), \quad  n_s \},   \label{9}
\end{equation}
\begin{equation}
\mathbf{P_6}=\{\Omega_bh^2, \quad \Omega_ch^2, \quad 100\theta_{MC}, \quad \tau, \quad \Sigma m_{\nu}, \quad  \mathrm{ln}(10^{10}A_s), \quad  n_s \},   \label{10}
\end{equation}
\begin{equation}
\mathbf{P_7}=\{\Omega_bh^2, \quad \Omega_ch^2, \quad 100\theta_{MC}, \quad \tau, \quad N_{eff}, \quad  \mathrm{ln}(10^{10}A_s), \quad  n_s \},   \label{10}
\end{equation}
where $\Omega_bh^2$ and $\Omega_ch^2$ are present-day baryon and CDM densities, $\theta_{MC}$ denotes the ratio between angular diameter distance and sound horizon at the redshift of last scattering $z_{ls}$, $\tau$ represents optical depth due to reionization, $\mathrm{ln}(10^{10}A_s)$ and $n_s$ are the amplitude and spectral index of primordial power spectrum at the pivot scale $K_0=0.05$ Mpc$^{-1}$, $\Omega_K$, $\omega$, $\beta$ and $\epsilon$ are typical parameters of o$\Lambda$CDM, $\omega$CDM, DEDP and DV models, respectively. Here $h$ is related to the Hubble constant $H_0$ by $H_0/h\equiv 100$ km s$^{-1}$ Mpc$^{-1}$.

The cosmological observations used in this analysis can be summarized in the following manner:

\textit{SNe Ia}: Testing the ability of the latest Pantheon SNe Ia sample in constraining the cosmological parameters is the key issue in this work. As described in the introduction, this sample integrates the PS1, SNLS, SDSS, low-$z$ and HST data to form the largest SNe Ia sample up to now, which consists of 1049 spectroscopically confirmed SNe Ia lying in the range $0.01<z<2.3$. D. Scolnic {\it et al.} have made a new progress on reducing the photometric calibration uncertainties, which dominates the systematic error budget of every major analysis of cosmological parameters with SNe Ia for a long time, to the point where they are familiar in magnitude to other major sources of known systematic errors (see \cite{11} for details). Hereafter we refer to this dataset as `` S ''.

\textit{CMB}: We employ the CMB temperature and polarization data from the full Planck-2015 survey in our numerical analysis \cite{17}, which includes the likelihoods of Planck-2015 low-$\ell$ temperature and polarization likelihood at $2\leqslant \ell\leqslant 29$, temperature (TT) at $30\leqslant \ell\leqslant 2500$, cross-correlation of temperature and polarization (TE) and polarization (EE) power spectra. Hereafter we denote this dataset as `` C ''.

\textit{BAO}: The BAO probe which is almost unaffected by uncertainties in the nonlinear evolution of matter density field and other systematic errors, is considered as the standard ruler to measure the evolution of the universe. To break the parameter degeneracies from other observations, here we use four BAO measurements: the 6dFGS sample at effective redshift $z_{eff}=0.106$ \cite{18}, the SDSS-MGS one at $z_{eff}=0.15$ \cite{19}, and the LOWZ at $z_{eff}=0.32$ and CMASS at $z_{eff}=0.57$ data from the SDSS-III BOSS DR12 sample \cite{20}. This dataset is denoted as `` B ''.

\textit{Cosmic chronometers}: We also employ the observations of cosmic expansion rate from cosmic chronometers (CC), which is independent of any cosmological theory. This dataset is obtained by utilizing the most massive and passively evolving galaxies based on the `` galaxy differential age ''. In this work, we use 30 CC data points to implement constraints on the above models \cite{21,22}. We shall refer to this dataset as `` H ''.

\textit{Lensing}: We also use the Planck-2015 lensing likelihood as a complementary probe to explore the evolution of the universe in this analysis \cite{23}. It is noteworthy that Planck-2015 lensing data has given the most powerful measurement with a 2.5$\%$ constraint on the amplitude of the lensing potential power
spectrum. We shall denote this dataset as `` L ''.

Using these observational datasets, we employ the Markov Chain Monte Carlo (MCMC) method to infer the posterior probability density distributions of different model parameters. Specifically, we modify carefully the November 2016 version of the publicly MCMC code \textbf{CosmoMC} \cite{24}, which obeys a
convergence diagnostic based on the Gelman and Rubin statistic, and Boltzmann code \textbf{CAMB} \cite{25}. To carry out the standard Bayesian analysis, we adopt the prior ranges for different model parameters as follows: $\Omega_bh^2 \in [0.005, 0.1]$, $\Omega_ch^2 \in [0.001, 0.99]$, $100\theta_{MC} \in [0.5, 10]$, $\tau \in [0.01, 0.8]$, $\mathrm{ln}(10^{10}A_s) \in [2, 4]$, $n_s \in [0.8, 1.2]$, $\Omega_K \in [-0.5, 0.5]$, $\omega \in [-3, 1]$, $\beta \in [-3, 3]$, $\epsilon \in [-0.3, 0.3]$, $\Sigma m_\nu \in [0, 5]$, $N_{eff} \in [3, 5]$. To exhibit better the ability of Pantheon sample in constraining seven cosmological models, firstly, we implement two kinds of constraints based on current data: (i) Pantheon alone (S); (ii) All data `` SCBHL ''.

\begin{table}[h!]
\renewcommand\arraystretch{1.3}

\caption{The 68$\%$ confidence intervals of different parameters of the $\Lambda$CDM, $o\Lambda$CDM, $\omega$CDM, DEDP and DV models using the combined datasets SCBHL, respectively.}
\label{t1}
\begin{tabular} { l c c c c c c}

\hline
\hline
Model & $\Lambda$CDM & $o\Lambda$CDM & $\omega$CDM & DEDP & DV \\
\hline
{$\Omega_b h^2   $} & $0.02241\pm 0.00013        $ & $0.02218\pm 0.00017 $ & $0.02233\pm 0.00015$ & $0.02238\pm 0.00012$  & $0.02244\pm 0.00014$                                                            \\

{$\Omega_c h^2   $} & $0.11756\pm 0.00077           $ & $0.11942^{+0.00067}_{-0.0013}        $ & $0.1182\pm 0.0011$ & $0.11766\pm 0.00070$ & $0.11779\pm 0.00083        $ \\

{$100\theta_{MC} $} & $1.04111\pm 0.00028$ & $1.04092\pm 0.00033       $ & $1.04097\pm 0.00030       $  & $1.04098\pm 0.00022 $ & $1.04106\pm 0.00029$ \\

{$\tau           $} &$0.0825^{+0.0017}_{-0.0027} $ & $0.0849^{+0.0026}_{-0.0032} $ & $0.078^{+0.012}_{-0.010}   $ & $0.0873^{+0.0051}_{-0.0046}   $ & $0.0825\pm 0.0031          $
                                                        \\
{${\rm{ln}}(10^{10} A_s)$} & $3.0967\pm 0.0035   $ & $3.1005^{+0.0029}_{-0.0037}  $ & $3.070^{+0.024}_{-0.027}   $  & $3.0971\pm 0.0060   $ & $3.084^{+0.0130}_{-0.0076}$                                                \\

{$n_s            $} & $0.9725\pm 0.0033          $ & $0.9643^{+0.0059}_{-0.0029}$ & $0.9688\pm 0.0043$ & $0.9707\pm 0.0030 $ & $0.9719^{+0.0032}_{-0.0028}       $                                                       \\

{$\Omega_K       $}  & ---  & $0.0020\pm 0.0014$ & ---  & ---  & ---                                                                                                            \\

{$\omega       $}  & ---  & --- & $-1.010^{+0.035}_{-0.031}$  & --- & ---                                                                                                           \\

{$\beta          $} & --- & --- &  ---  & $-0.0026\pm 0.0017$ & ---                                                    \\

{$\epsilon          $} & --- & --- &  --- & --- & $-0.00029^{+0.00028}_{-0.00025}        $                                                     \\

\hline
$H_0              $ & $68.31\pm 0.37    $ & $68.41^{+0.88}_{-0.53}            $ & $68.26^{+0.75}_{-0.87}       $ & $68.21\pm 0.28        $ & $68.33\pm 0.37      $                                                      \\
$\Omega_m              $ & $0.3013\pm 0.0046$ & $0.3040^{+0.0049}_{-0.0090}          $ & $0.3031\pm 0.0074  $ & $0.3024\pm 0.0038   $ & $0.3018\pm 0.0048       $                                             \\
$\sigma_8              $ & $0.8266\pm 0.0031   $ & $0.8341^{+0.0035}_{-0.0053}$ & $0.8204^{+0.0102}_{-0.0122}   $ & $0.8265\pm 0.0039    $ & $0.8219\pm 0.0052           $                                          \\
\hline
$\chi^2_{min}$ & 14051.17 & 14053.99 & 14051.68 & 14052.66 & 14049.42 \\
\hline
\hline
\end{tabular}
\end{table}

\begin{table}[h!]
\renewcommand\arraystretch{1.3}

\caption{The 68$\%$ confidence intervals of different parameters of the $o\Lambda$CDM, $\omega$CDM, DEDP and DV models using the Pantheon SNe Ia sample, respectively. }
\label{t2}
\begin{tabular} { l c c c c c c}

\hline
\hline
Model & $\Lambda$CDM & $o\Lambda$CDM & $\omega$CDM & DEDP & DV \\
\hline

$\Omega_m         $    & $0.297\pm 0.024           $   & $0.318\pm 0.071           $ & $0.312^{+0.082}_{-0.058} $ & $0.305\pm 0.017    $ & $0.302\pm 0.021      $                                             \\

{$\Omega_K       $}     & ---      & $-0.058\pm0.121$ & ---  & ---  & ---                                                                                                            \\

{$\omega       $} & ---    & --- & $-1.07^{+0.22}_{-0.20}$  & --- & ---                                                                                                           \\

{$\beta          $} & ---  & --- &  ---  & $0.10^{+0.12}_{-0.21} $ & ---                                                    \\

{$\epsilon          $} & ---  & --- &  --- & --- & $0.0079^{+0.0099}_{-0.0077}        $                                                     \\

\hline
$\chi^2_{min}$ &1036.48 & 1036.74 & 1037.85 & 1036.64 & 1036.90  \\
\hline
\hline
\end{tabular}
\end{table}

\begin{figure}
\centering
\includegraphics[scale=0.45]{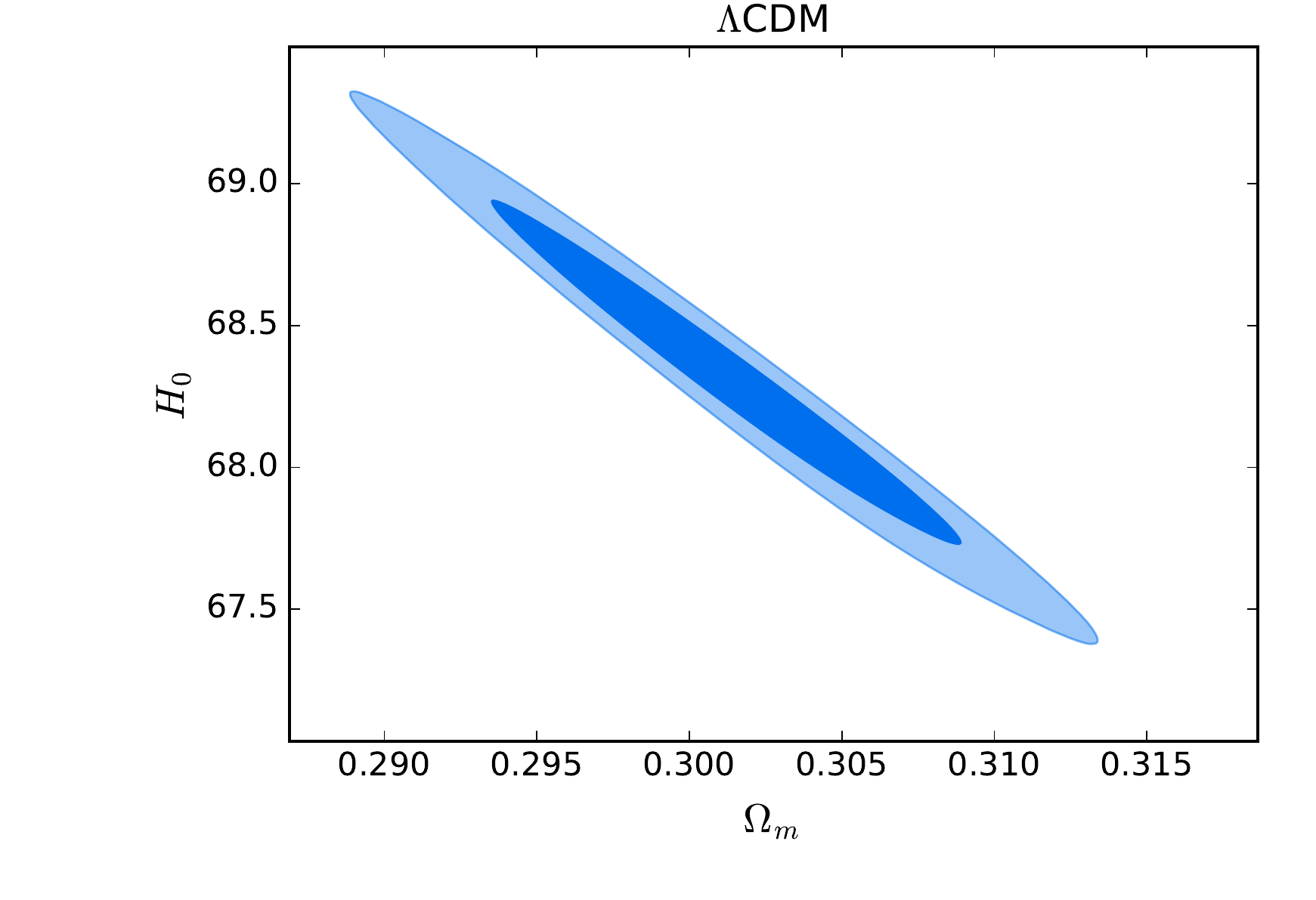}
\caption{The $68\%$ and $95\%$ confidence regions of the $\Lambda$CDM model using the combined datasets SCBHL are shown.}\label{f1}
\end{figure}

\begin{figure}
\centering
\includegraphics[scale=0.49]{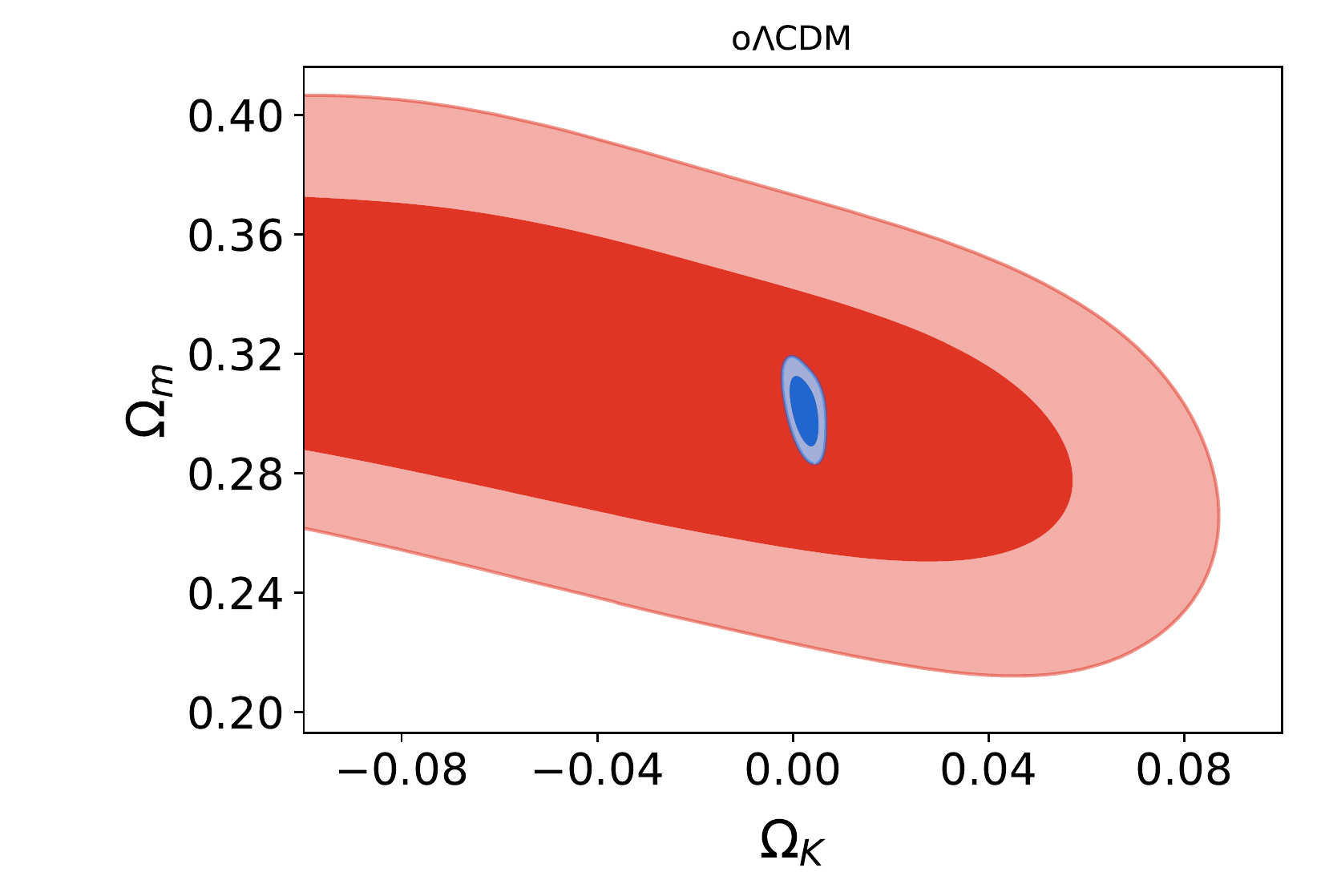}
\includegraphics[scale=0.45]{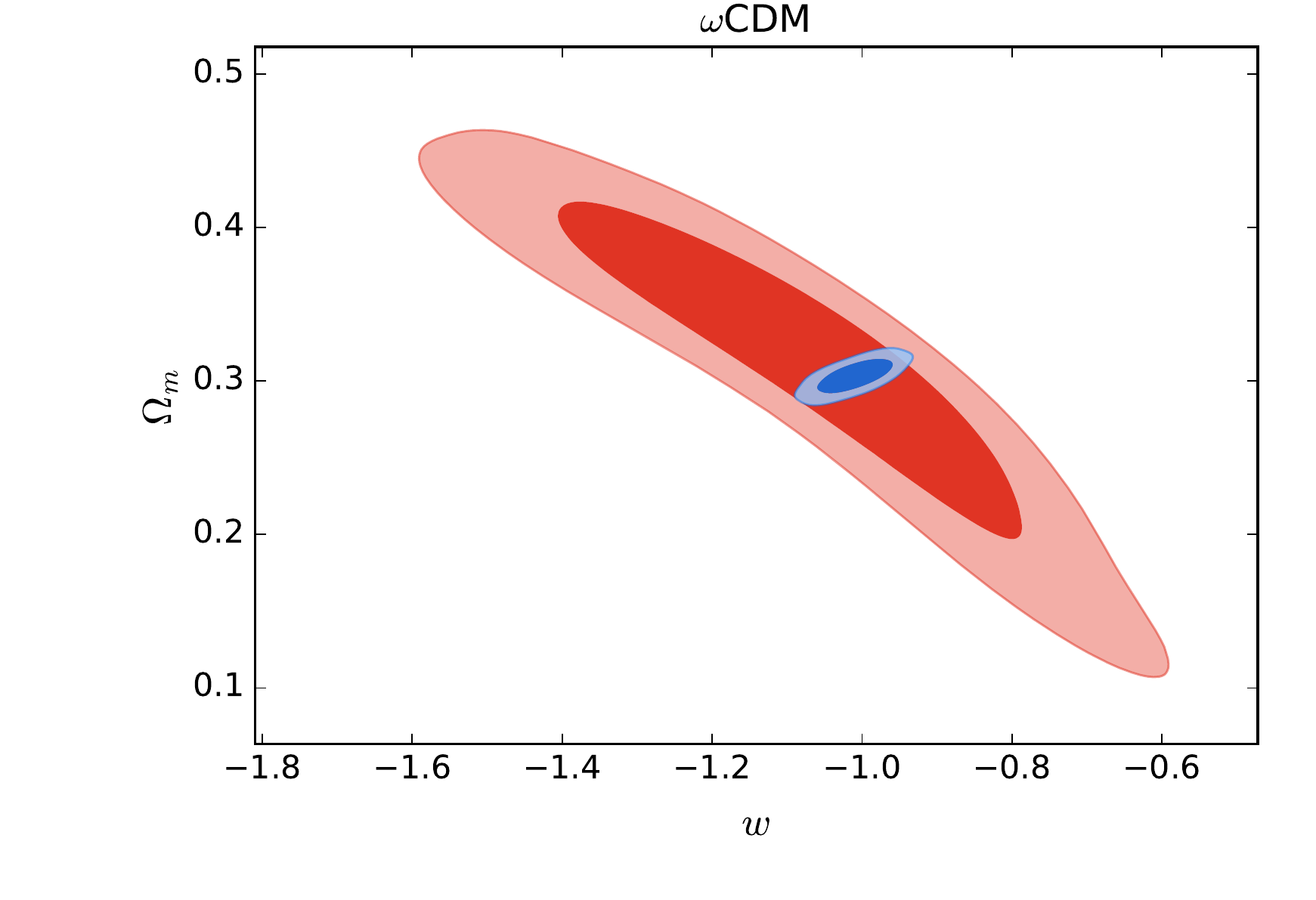}
\includegraphics[scale=0.45]{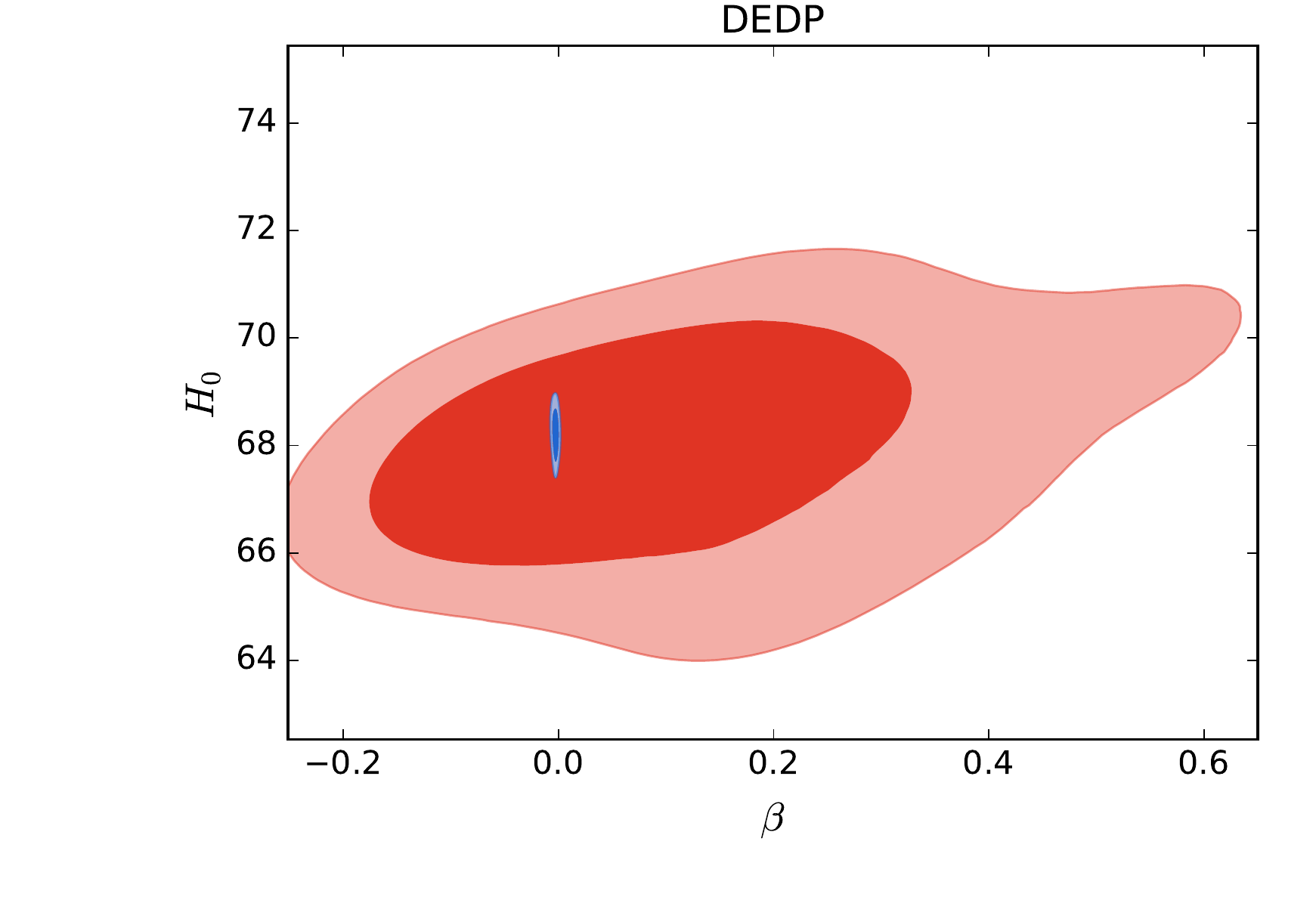}
\includegraphics[scale=0.45]{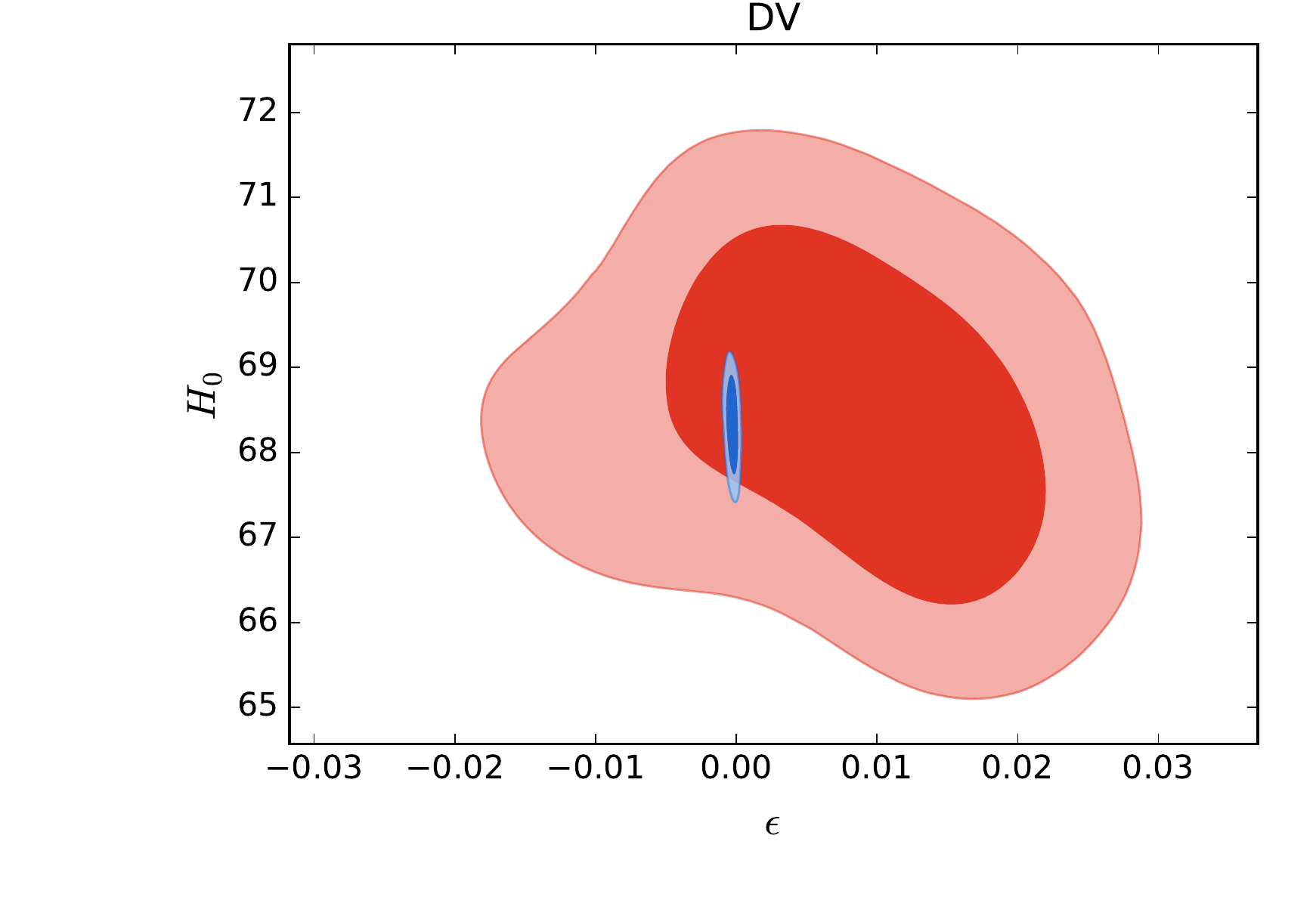}
\caption{The $68\%$ and $95\%$ confidence regions of the $o\Lambda$CDM, $\omega$CDM, DEDP and DV models using the Pantheon SNe Ia sample (red contours) and combined datasets SCBHL (blue contours) are shown, respectively.}\label{f2}
\end{figure}

\begin{figure}
\centering
\includegraphics[scale=0.45]{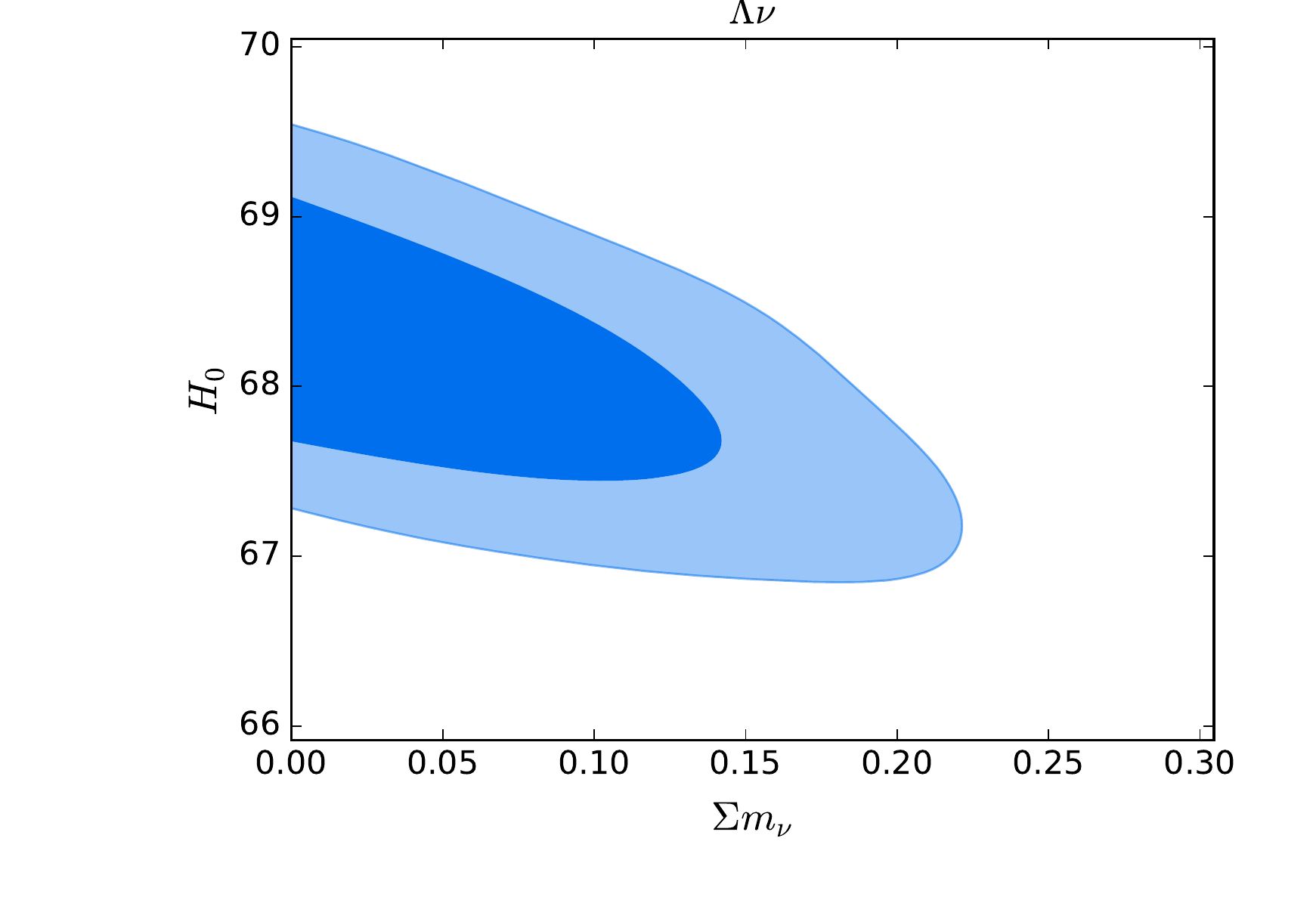}
\includegraphics[scale=0.45]{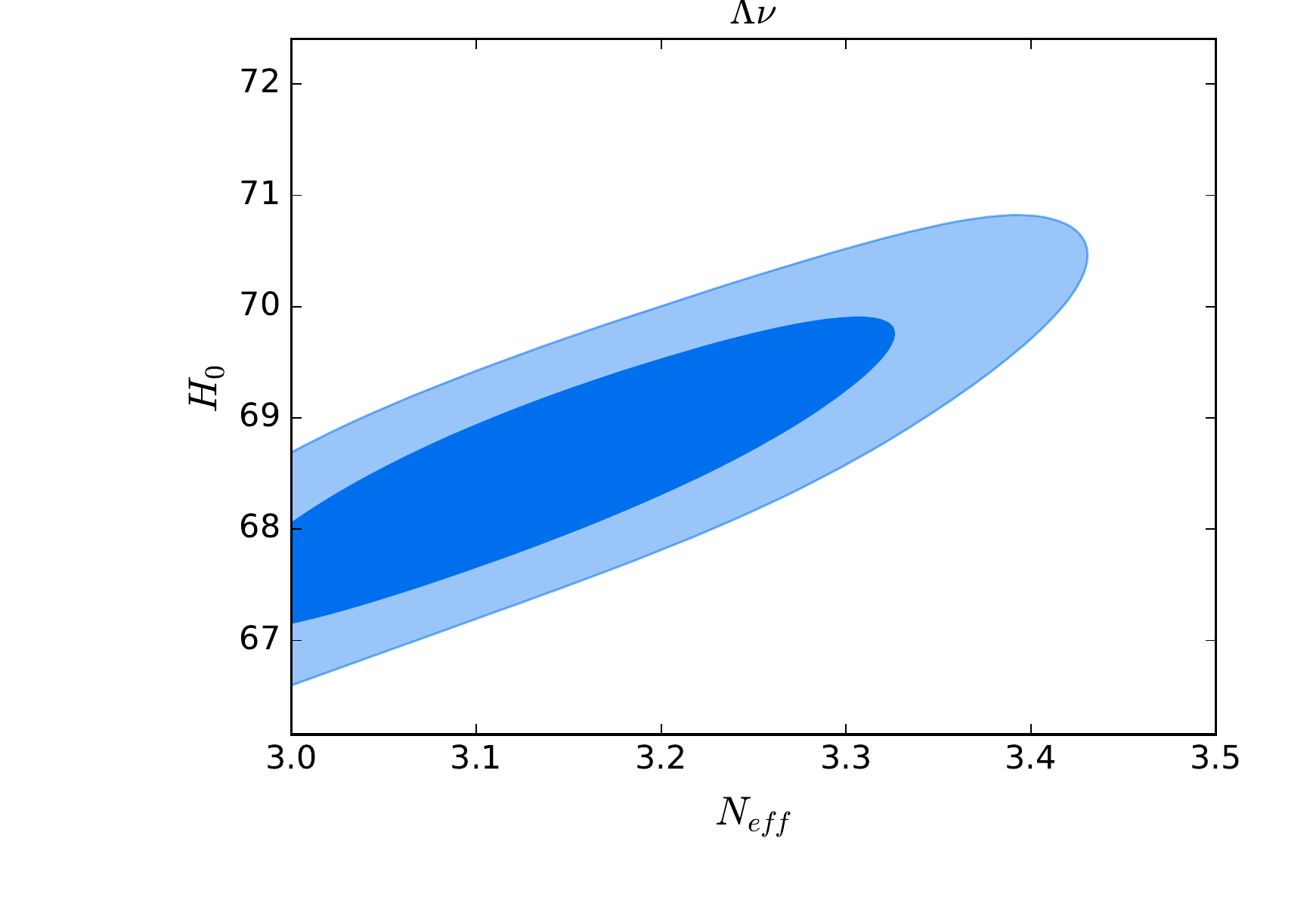}
\caption{The $68\%$ and $95\%$ confidence regions of the $\Lambda\nu$ and $\Lambda$s models using the combined datasets SCBHL are shown, respectively.}\label{f3}
\end{figure}

\section{analysis results}
Utilizing the combined constraint from SCBHL, our MCMC results are exhibited in Tab. \ref{t1}. The 68$\%$ and 95$\%$ confidence regions of key parameter pairs of the $\Lambda$CDM, $o\Lambda$CDM, $\omega$CDM, DEDP, DV, $\Lambda\nu$ and $\Lambda$s models are also presented in Figs. \ref{f1}- \ref{f3}, respectively. Combining the CBHL datasets with the Pantheon sample, it is very clear that the constraints on the parameters of $\Lambda$CDM, $o\Lambda$CDM and $\omega$CDM models are improved in comparison with the results shown in \cite{11} (see Tab. \ref{t1}). More specifically, using SCBHL, we obtain the following conclusions: (i) for $o\Lambda$CDM, the spatial curvature $\Omega_K=0.0020\pm0.0014$ of the universe is compatible with zero at the 1.43$\sigma$ confidence level (CL), which indicates that a spatially flat universe is preferred by current data in the framework of $\Lambda$CDM cosmology. Meanwhile, our result prefers a positive best-fitting value corresponding to an open universe and is in a good agreement with the Planck's restriction $|\Omega_K | < 0.005$ \cite{17}; (ii) for $\omega$CDM, the constrained EoS of DE $\omega=-1.010^{+0.035}_{-0.031}$ is consistent with the cosmological constant scenario and Planck's result $\omega=-1.006\pm0.045$ at the 1$\sigma$ CL \cite{17}; (iii) for DEDP, the typical model parameter $\beta=-0.0026\pm 0.0017$ is consistent with zero at the 1.53$\sigma$ CL, which is very compatible with our previous result \cite{12} and implies that there is no obvious evidence for dynamical DE; (iv) for DV, the constrained modified matter expansion rate $\epsilon=-0.00029^{+0.00028}_{-0.00025}$ just deviates very slightly from zero, which indicates that there is no obvious interaction between DM and DE in the dark sector of the universe at the 1.04$\sigma$ CL; (v) for $\Lambda\nu$, the mass sum of three active neutrinos is constrained to $\Sigma m_\nu<0.178$ eV at the 2$\sigma$ CL, which is tighter than the Planck's constraint $\Sigma m_\nu<0.23$ eV by using `` Planck TT + lowP + lensing + ext '' (see \cite{17} for more details); (vi) for $\Lambda$s, we give the $68\%$ limits of the effective number of relativistic degrees of freedom $N_{eff}=3.11\pm0.15$, which is compatible with the standard value $N_{eff}=3.046$ of the Standard Model of particle physics at the 1$\sigma$ CL and has a smaller error than Planck's restriction $N_{eff}=3.15\pm0.23$ \cite{17}. In Fig. \ref{f2}, one can also find that the constraints on the typical parameters $\Omega_K$, $\omega$, $\beta$ and $\epsilon$ from SCBHL are much tighter than those from the Pantheon sample. By combining the Pantheon dataset with CBHL, we find that the current $H_0$ tension between the directly local observation $H_0=73.24\pm1.74$ km s$^{-1}$ Mpc$^{-1}$ by Riess {\it et al.} \cite{26} and the indirectly global derivation $H_0=66.93\pm0.62$ km s$^{-1}$ Mpc$^{-1}$ by Planck Collaboration \cite{17} can be slightly alleviated from 3.4$\sigma$ to 2.77$\sigma$, 2.48$\sigma$, 2.63$\sigma$, 2.85$\sigma$, 2.76$\sigma$, 2.85$\sigma$ and 2.40$\sigma$ for the $\Lambda$CDM, $o\Lambda$CDM, $\omega$CDM, DEDP, DV, $\Lambda\nu$ and $\Lambda$s models, respectively. Meanwhile, as Planck's prediction $n_s=0.968\pm0.006$ \cite{17}, we find that the scale invariance of primordial power spectrum in the $\Lambda$CDM, $o\Lambda$CDM, $\omega$CDM, DEDP, DV, $\Lambda\nu$ and $\Lambda$s models is strongly excluded at the 8.33$\sigma$, 6.05$\sigma$, 7.26$\sigma$, 9.77$\sigma$, 8.33$\sigma$, 7.34$\sigma$ and 5.15$\sigma$ CL, respectively.

We also exhibit the MCMC results using the Pantheon sample to constrain cosmological models in Tab. \ref{t2}.  We find that: (i) our constraints $\Omega_m=0.297\pm 0.024$ for $\Lambda$CDM, $\Omega_K=-0.058\pm0.121$ and $\Omega_m=0.318\pm 0.071$ for $o\Lambda$CDM, and $\omega=-1.07^{+0.22}_{-0.20}$ and $\Omega_m=0.312^{+0.082}_{-0.058}$ for $\omega$CDM are well consistent with those presented in \cite{11}; (ii) using only Pantheon data, the constrained parameters $\beta=0.10^{+0.12}_{-0.21}$ for DEDP and $\epsilon=0.0079^{+0.0099}_{-0.0077}$ for DV imply that there is still no evidence of dynamical DE at the 1$\sigma$ CL and no interaction between DM and DE in the dark sector of the universe at the 1.03$\sigma$ CL, respectively; (iii) the constraints $\omega=-1.010^{+0.035}_{-0.031}$ for $\omega$CDM and $\epsilon=-0.00029^{+0.00028}_{-0.00025}$ for DV, and $\beta=-0.0026\pm 0.0017$ for DEDP using SCBHL are more accurate than those using only Pantheon sample by one and two orders of magnitudes, respectively; (iv) very interestingly, for $o\Lambda$CDM, our result $\Omega_K=-0.058\pm0.121$ from SNe Ia data prefers a negative best-fitting value corresponding to a closed universe, which is different from the case of SCBHL (see also Tabs. \ref{t1}-\ref{t2}).

\section{Discussions and Conclusions}
Based on the most recent work that releases the largest Pantheon SNe Ia sample to date \cite{11}, our motivation is to test the ability of the Pantheon SNe Ia in constraining cosmological models, implement strict constraints using the data combination SCBHL and investigate whether there is new physics beyond the standard cosmological model.

Using the combined datesets SCBHL, we find that: (i) a spatially flat universe is supported in the framework of $\Lambda$CDM cosmology; (ii) the constrained EoS of DE is very consistent with the cosmological constant scenario at the $1\sigma$ CL in the $\omega$CDM model; (iii) there does not exist the dynamical DE by constraining the DEDP model we recently proposed; (iv) there is no interaction between DM and DE in the dark sector of the universe in the DV model; (v) there is no obvious hint of sterile neutrinos in the neutrino sector of the universe under the assumption of $\omega$CDM model; (vi) the $H_0$ tension between the directly local observation by Riess {\it et al.} \cite{26} and the indirectly global derivation by Planck Collaboration \cite{17} can be moderately relieved in all seven models.

It is noteworthy that we obtain improved constraints on neutrino parameters $\Sigma m_\nu$ and $N_{eff}$ by adding two updated BAO data points from BOSS DR12 sample \cite{20}, CMB TT and EE likelihoods and cosmic chronometers into the Planck's analysis \cite{17}. The effects of the improvements can be mainly attributed to the addition of two updated BAO points and CMB TT and EE likelihoods. Other than the data used in \cite{11}, we switch one old BAO point for two newer BAO ones, and also use CMB TT and EE likelihoods, Planck lensing data and cosmic chronometers to constrain the above cosmological models and obtain our better constraints on different cosmological parameters.

Since the constraint on the mass sum of three active neutrinos $\Sigma m_\nu<0.178$ eV we give is tighter than the mass range $\Sigma m_\nu<0.23$ eV preferred by the Planck Collaboration \cite{17}, the direct detection of non-relativistic cosmic neutrinos with first generation experiment such as the PTOLEMY would be very difficult \cite{27,28,29}.

It is clear that we do not find any departure from the $\Lambda$CDM model based on current data. However, in theory, there is still a lack of deep understandings about the nature of DE. In the future, we hope to indicate the theoretical direction from the view of cosmological data analysis.

\section{acknowledgements}
We are grateful to the referee for helping us improve this work.
We thank Yang-Jie Yan, Wei Zhang, Jing-Ling Chen and Xin-He Meng for very useful communications for a long time.  We also thank Liu Zhao, Yi Liao, Fu-Lin Zhang, Da-Bao Yang, Lei Fang and Yuan Sun for helpful discussions on cosmology, gravity, particle physics and quantum information.


\begin{thebibliography}{99}
\bibitem{1}
A. G. Riess {\it et al.} [Supernova Search Team], Astron. J. {\bf 116}, 1009 (1998).

\bibitem{2}
S. Perlmutter {\it et al.} [Supernova Cosmology Project], Phys. Rev. Lett. {\bf 83}, 670 (1999).

\bibitem{3}
D. H. Weinberg {\it et al.}, Phys. Rep. {\bf 530}, 87 (2013).


\bibitem{4}
C. L. Bennett {\it et al.} [WMAP Collaboration], Astrophys. J. Suppl. Ser. {\bf 208}, 20 (2013).

\bibitem{5}
P. Ade {\it et al.} [Planck Collaboration], Astron. Astrophys. {\bf 571}, A16 (2014).

\bibitem{6}
  B.~P.~Abbott {\it et al.} [LIGO Scientific and Virgo Collaborations],
  Phys.\ Rev.\ Lett.\  {\bf 119}, 161101 (2017).

\bibitem{7}
  B.~P.~Abbott {\it et al.} [LIGO Scientific and Virgo and 1M2H and Dark Energy Camera GW-E and DES and DLT40 and Las Cumbres Observatory and VINROUGE and MASTER Collaborations],
  Nature {\bf 551}, 85 (2017).

\bibitem{8}
  S.~M.~Carroll,
  Living Rev.\ Rel.\  {\bf 4}, 1 (2001).

\bibitem{9}
S. Weinberg, Rev. Mod. Phy. {\bf 61}, 1 (1989).

\bibitem{10}
M. Betoule {\it et al.} [SDSS collaboration], Astron. Astrophys. {\bf 568},  A22 (2014).

\bibitem{11}
  D.~M.~Scolnic {\it et al.},
  arXiv:1710.00845 [astro-ph.CO].

\bibitem{12}
  D.~Wang and X.~Meng,
  Phys.\ Rev.\ D {\bf 96}, 103516 (2017).

\bibitem{13}
 P.~Wang and X.~Meng,, Class. Quantum Grav. {\bf 22}, 283 (2005).

\bibitem{a1}
  D.~Wang and X.~Meng,
  Phys.\ Rev.\ D {\bf 96}, 023538 (2017).

\bibitem{a2}
  D.~Wang and X.~Meng,
  Eur.\ Phys.\ J.\ C {\bf 77}, 725 (2017).

\bibitem{a3}
  D.~Wang and X.~Meng,
  arXiv:1709.04141 [astro-ph.CO].

\bibitem{14}
  V.~F.~Mukhanov, H.~A.~Feldman and R.~H.~Brandenberger,
  Phys.\ Rept.\  {\bf 215}, 203 (1992).

\bibitem{15}
  C.~P.~Ma and E.~Bertschinger,
  Astrophys.\ J.\  {\bf 455}, 7 (1995).

\bibitem{16}
  K.~A.~Malik and D.~Wands,
  Phys.\ Rept.\  {\bf 475}, 1 (2009).

\bibitem{17}
P. Ade {\it et al.} [Planck Collaboration], Astron. Astrophys. {\bf 594}, A13 (2016).

\bibitem{18}
F. Beutler {\it et al.}, Mon. Not. Roy. Astron. Soc. {\bf 3017}, 416 (2011).

\bibitem{19}
A. J. Ross {\it et al.}, Mon. Not. Roy. Astron. Soc. {\bf 835}, 449 (2015).

\bibitem{20}
A. J. Cuesta {\it et al.},  Mon. Not. Roy. Astron. Soc. {\bf 457}, 1770 (2016).

\bibitem{21}
M. Moresco {\it et al.},  J. Cosmol. Astropart. Phys. {\bf 05} (2016) 014.

\bibitem{22}
D. Wang, X. Meng, Phys. Rev. D {\bf 95}, 023508 (2017).

\bibitem{23}
P. Ade {\it et al.} [Planck Collaboration], Astron. Astrophys. {\bf 594}, A15 (2016).

\bibitem{24}
A. Lewis, Phys. Rev. D {\bf 87}, 103529 (2013).

\bibitem{25}
A. Lewis, A. Challinor, and A. Lasenby, Astrophys. J. {\bf 538}, 473 (2000).

\bibitem{26}
A. G. Riess {\it et al.}, Astrophys. J. {\bf 826}, 56 (2016).

\bibitem{27}
  A.~G.~Cocco, G.~Mangano and M.~Messina,
  J. Cosmol. Astropart. Phys. {\bf 0706}, 015 (2007).

\bibitem{28}
  S.~Betts {\it et al.},
  arXiv:1307.4738 [astro-ph.IM].

\bibitem{29}
  A.~J.~Long, C.~Lunardini and E.~Sabancilar,
  J. Cosmol. Astropart. Phys. {\bf 1408}, 038 (2014).



























































































\end{thebibliography}
\end{document}